\documentclass[acmsmall, review]{acmart}

\usepackage{graphicx,balance,multirow,xspace,subcaption,longtable}
\usepackage{tabularx,multirow,booktabs,blindtext}
\usepackage{makecell,amsmath,cleveref,rotating}
\usepackage[font={small}, textfont=md]{caption}
\usepackage[T1]{fontenc}
\usepackage[utf8]{inputenc}
\usepackage[draft,inline,nomargin,index]{fixme}
\usepackage{marginnote}
\usepackage{booktabs}
\usepackage{xcolor}

\setcopyright{none}

\AtBeginDocument{%
  }

\PassOptionsToPackage{hyphens}{url}
\usepackage{hyperref} 
\newcolumntype{P}[1]{>{\arraybackslash}p{#1}}
\newcolumntype{X}[1]{>{\centering\arraybackslash}p{#1}}

\expandafter\def\expandafter\UrlBreaks\expandafter{\UrlBreaks
  \do\a\do\b\do\c\do\d\do\e\do\f\do\g\do\h\do\i\do\j%
  \do\k\do\l\do\m\do\n\do\o\do\p\do\q\do\r\do\s\do\t%
  \do\u\do\v\do\w\do\x\do\y\do\z\do\A\do\B\do\C\do\D%
  \do\E\do\F\do\G\do\H\do\I\do\J\do\K\do\L\do\M\do\N%
  \do\O\do\P\do\Q\do\R\do\S\do\T\do\U\do\V\do\W\do\X%
  \do\Y\do\Z}

\crefformat{section}{\S#2#1#3}
\crefformat{subsection}{\S#2#1#3}
\crefformat{subsubsection}{\S#2#1#3}

\hypersetup{%
  pdftitle={Google Search},
  pdfauthor={},
  pdfkeywords={},
  bookmarksnumbered,
  pdfstartview={FitH},
  colorlinks,
  citecolor=black,
  filecolor=black,
  linkcolor=black,
  urlcolor=linkColor,
  breaklinks=true,
  hypertexnames=false
}

\newcommand\clearrow{\global\let\rowmac\relax}

\newcommand{\para}[1]{{\vspace{.05in} \bf \noindent #1 }}

\copyrightyear{2019} 
\acmYear{2018} 
\setcopyright{acmlicensed}
\acmConference[IMC '19]{acm}{imc}

\setcopyright{none}

\begin{document}

\renewcommand\footnotetextcopyrightpermission[1]{} 
\pagestyle{plain} 

\renewcommand{\sectionautorefname}{\S}
\renewcommand{\subsectionautorefname}{\S}
\renewcommand{\subsubsectionautorefname}{\S}

\title{Algorithmic Amplification of biases on Google Search}
\subtitle{}

\author{Hussam Habib}
\affiliation{%
  \institution{University of Iowa}
  \city{Iowa City}
  \country{USA}}
\email{hussam-habib@uiowa.edu}

\author{Ryan Stoldt}
\affiliation{%
  \institution{Drake University}
  \country{USA}}

\author{Andrew High}
\affiliation{%
  \institution{Pennsylvania State University}
  \country{USA}}

\author{Brian Ekdale}
\affiliation{%
  \institution{University of Iowa}
  \country{USA}}

\author{Ashley Peterson}
\affiliation{%
  \institution{Peloton Interactive}
  \country{USA}}

\author{Katy Biddle}
\affiliation{%
  \institution{University of Iowa}
  \country{USA}}

\author{Javie Ssozi}
\affiliation{%
  \institution{University of Iowa}
  \country{USA}}

\author{Rishab Nithyanand}
\affiliation{%
  \institution{University of Iowa}
  \city{Iowa City}
  \country{USA}}
\email{rishab-nithyanand@uiowa.edu}

\renewcommand{\shortauthors}{Habib et al.}
\begin{abstract}
The evolution of information-seeking processes, driven by search engines like
Google, has transformed the access to information people have. This paper
investigates how individuals' pre-existing attitudes influence the modern
information-seeking process, specifically, the results presented by Google
Search. Through a comprehensive study involving surveys and information-seeking
tasks focusing on the topic of abortion, the paper provides four crucial
insights: 1) Individuals with opposing attitudes on abortion receive different
search results. 2) Individuals express their beliefs in their choice of
vocabulary used in formulating the search queries, shaping the outcome of the
search. 3) Additionally, the user’s search history contributes to divergent
results among those with opposing attitudes. 4) Google Search engine reinforces
pre-existing beliefs in search results. Overall, this study provides insights
into the interplay between human biases and algorithmic processes, highlighting
the potential for information polarization in modern information-seeking
processes.
\end{abstract}



\maketitle
\sloppy
\section{Introduction}
Processes for seeking and consuming information have changed significantly over
the past two decades. Search engines like Google now play a vital role in
information-seeking by allowing people to easily find information online by
algorithmically sorting information and personalizing the information based on
user data. When faced with an information-seeking task, users frequently begin
by formulating a query to be input into a search engine, most commonly Google
\cite{SearchEngineMarket}. Once the query is entered into the search engine,
several complex, opaque, and ever-changing computational algorithms index,
interpret, filter, and deliver related information to the user
\cite{GoogleSearchStatus}.

The entire algorithmic information-seeking process diverges in many ways from
traditional information seeking with some subprocesses introducing undesirable
patterns. One such pattern involves the formation of filter bubbles, which occur
when algorithms disproportionately curate content that reinforces users'
preexisting beliefs \cite{pariserFilterBubbleWhat2011}. Filter bubbles provide
pathways for the formation of “alternate realities” and can contribute to the
formation of extremist ideologies.

This paper explores how individuals' preexisting attitudes on a topic influence,
explicitly or implicitly, the information made available by search engine
algorithms. Our study focuses on the information-seeking behavior of Google users
on the contentious topic of abortion. Participants were surveyed about their
attitudes on abortion and then prompted to engage in multiple
information-seeking tasks specifically related to abortion. They were instructed
to formulate queries based on these prompts, perform Google Searches
accordingly, and subsequently share the search results they encountered. We
analyze the data to answer each of the following questions and present a
comprehensive picture of human and algorithm interplay in the modern
information-seeking process:

\begin{itemize}
  \item \textit{RQ1: Do individuals with opposing attitudes on a given topic receive
  different search results?} In other words, during an information-seeking
  process on a topic, do the individual's preexisting attitudes on the topic
  influence the search results they are presented with? Given participants with
  varying attitudes towards abortion and their formulated queries, we compare
  search results presented to participants with opposite preexisting attitudes,
  i.e., pro-life, and pro-choice. We find there to be significant differences in
  the domains, titles, and snippets presented.
  \item \textit{RQ2: Do individuals encode their preexisting beliefs within
  their search queries, subsequently influencing the outcomes they obtain?} In
  this section, we investigate whether individuals formulate queries for their
  information task based on their preexisting attitudes on the topic,
  subsequently influencing divergent search results. To this end, we compare the
  queries formulated by participants with opposing attitudes on abortion. We
  find the preexisting attitudes do not influence the semantics or style of the
  queries, however, the choice of words within the queries across opposing
  groups was significantly different. This nuanced difference in vocabulary
  choice based on the participant's preexisting attitudes seems to influence the
  search results, as demonstrated by our mediation model.
  \item \textit{RQ3: Does search history contribute to the divergence of results
  among users holding opposing attitudes?} In addition to queries, we
  investigate the role of search history and personalization in influencing
  search results. We measure the influence of pre-existing abortion attitudes on
  search results obtained through a controlled lab experiment. By repeating the
  searches using queries formulated by the participants in a controlled lab
  computer we remove any influence of search history on the divergent search
  results based on preexisting attitudes. We find search results to remain
  influenced indirectly by the abortion attitudes of participants through their
  formulated queries alone. 
  \item \textit{RQ4: To what extent does Google serve results that reinforce
  preexisting attitudes?} Our prior findings established the communication of
  preexisting beliefs to the search engine and its influence to present
  diverging search results. Finally, we investigate the meaningful
  characteristics in which the search results are different across participants
  with opposing attitudes to determine a filter bubble effect. Comparing the
  sources, we find participants with a pro-life stance towards abortion are more
  likely to be presented with search results from right-biased news sources.
  Whereas, analyzing the titles and snippets of the search results semantically,
  participants were more likely to be shown search results that used vocabulary
  associated with their preexisting attitudes. 
\end{itemize}

\begin{figure}[!htbp]
  \centering
  \includegraphics[width=\linewidth]{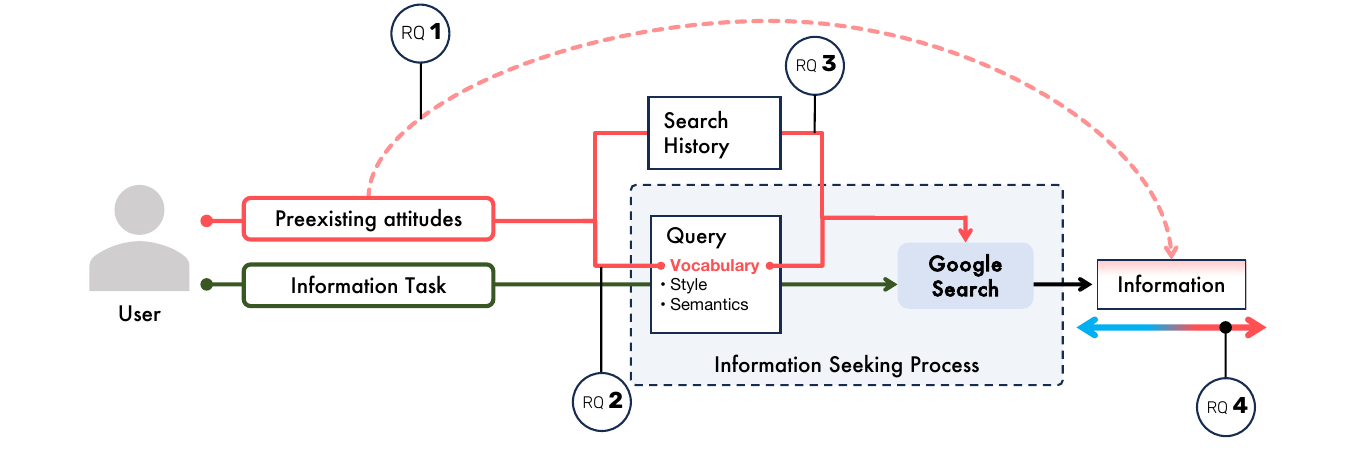}
  \label{fig:ISP}
  \caption{The modern information-seeking process. We investigate the following
  relationships:
  \newline
  RQ1. Do preexisting attitudes influence search results?  
  \newline
  RQ2. How do formulation of queries mediate the influence of attitudes on
  results? 
  \newline
  RQ3. How do search history and personalization influence search results?
  \newline
  RQ4. Do the results reinforce preexisting attitudes?
  }
\end{figure}

Taken together, we synthesize prior conclusions about fragments of the
information-seeking process and address four questions that investigate the
relationship between algorithms and cognitive biases on the Google search
platform. 

\section{Background}
\label{sec:background}
Prior studies have investigated the information-seeking process at various
stages to uncover how search engines and their algorithms influence the
information users ultimately acquire. This research has shown that users tend to
encode their predispositions in search queries, even to the point of revealing
their political beliefs \cite{vanhoofSearchingDifferentlyHow2022,
slechtenAdaptingSelectiveExposure2022, PropagandistsPlaybook}. These findings suggest the presence of
cognitive biases during query formulation, which in effect facilitates the
self-selection of information and increases the potential for filter bubbles to
occur. Focusing on political news information seeking and consumption, strong
evidence for the formation of filter bubbles has been identified \cite{leMeasuringPoliticalPersonalization2019a}. Despite
these findings, recent studies have also made conflicting discoveries regarding
the occurrence of filter bubbles. Notably, research relying on artificial agents
\cite{haimAbyssShelterRelevance2017, hannakMeasuringPersonalizationWeb2017} and
real users \cite{ochigameSearchAtlasVisualizing2021, robertsonAuditingPartisanAudience2018} to generate a variety of user profiles found that
algorithmic personalization did not result in the formation of filter bubbles.
These findings were also supported by research explicitly focused on the effects
of algorithmic filtering \cite{puschmannBubbleAssessingDiversity2019}.
Crucially, Slechten et al. \cite{slechtenAdaptingSelectiveExposure2022} found
users to not only be more likely to write queries confirming their preexisting
beliefs but also, when confronted with search results that do not align with their prior
beliefs the participants were more likely to try again. However, surprisingly,
investigating the influence of algorithmic intervention in seeking information,
the authors find that search results that were ranked higher were more likely to
be engaged with. These search results were also more likely to be neutral in
nature, as opposed to search results that confirmed or disconfirmed the user's
prior beliefs. These findings suggest that the information-seeking process is a
complex interplay between the user's preexisting beliefs and the algorithmic
intervention.
Taken together, these studies, each studying one fragment of the modern
information-seeking process, present a conflicting picture of the dynamics
between users and algorithms. Ultimately, it is yet unclear whether search
engines construct filter bubbles and, if filter bubbles exist, if this effect is
caused by human cognitive biases or algorithmic behaviors. These gaps in
knowledge beg for a comprehensive analysis of the complete information-seeking
process, which this paper aims to provide. 

\section{Data}
\label{sec:data}

\subsection{Survey}
\label{sec:data:survey}
We fielded our survey (N=227) in June 2022 at a large university on the East
Coast. Respondents identified as primarily male (66.1\%) over female (32.6\%),
with two identifying as transgender and one preferring not to answer. Regarding
race and ethnicity, .9\% identified as African American, 5\% as Asian/Pacific
Islander, .9\% Latinx, 74.8\% White/European American, 5\% Other, 6.4\% Asian
American, and 7\% as Multiple races. The majority of respondents' ages ranged
from 18-20 (96.3\%) with 2.3\% between 21-29 and 1.5\% over 29 years old. After
obtaining Institutional Review Board (IRB) approval, our research team fielded
the survey to undergraduate participants, who were told that the purpose of the
survey was “to better understand how people search for current events and other
topics.” 

\subsection{Abortion Scale}
\label{sec:data:abortion-scale}
To measure our participant's views on abortion, we adapted a previously
validated scale  \cite{taylorMeasurementAttitudesAbortion2014} by Hess and Reub
\cite{hessAttitudesAbortionReligion2005}. The scale is comprised of nine items ,
seven of which came from Hess \& Reub. In light of recent anti-abortion bills
introduced across various states, we added two original items to the scale: “I
support laws requiring women to wait 24 hours between meeting with a healthcare
provider and getting an abortion.” and “I support laws requiring doctors to show
and describe ultrasound images to women seeking an abortion".


All items in the scale were measured on a Likert scale from one to
seven. For the aggregate scale, high scores correlate with being pro-life, while
low scores correlate with being pro-choice. To confirm the fit of our adapted
scale to the data, we performed a confirmatory factor analysis. An initial model
in which all nine items loaded on a latent factor produced poor fit, $X^2$= 392.09,
$p$ < .001, $CFI$ = .69, $RMSEA$ = .245, $SRMR$ = .12. Consulting modification indices
suggested that we correlate three pairs of error terms, each of which seemed
logical (i.e., reverse-coded items, same topical focus). We iteratively re-ran
the model, and after correlating those three pairs of error terms, we found a
model that fit the data, $X^2$= 66.54, $p$ < .001, $CFI$ = .96, $RMSEA$ = .09, $SRMR$ = .07.
In our study, participants were categorized into three groups using the
aggregate scale: neutral, pro-life, and pro-choice. We consider participants
with scores on the aggregate abortion scale 3 \& below as pro-choice, 5 \& above
as pro-choice, and between 3 and 4 to be neutral.

\subsection{Queries}
\label{sec:data:queries}
Following our collection of participants' stances on abortion, we prompt the
participants to write queries related to abortion. We ask our participants to
write a query each for eight prompts we provide (see \Cref{tab:prompts}). Each prompt was
designed to simulate a scenario where the participant comes across a post on
social media related to abortion. The eight prompts included four closed-ended
prompts and four open-ended prompts. Closed-ended prompts were designed to
trigger fact-finding searches, in which the searcher seeks out a single,
fact-based answer, while open-ended prompts were designed to trigger
information-gathering searches, in which the searcher would engage in a broader
effort to collect information from a variety of sources
\cite{roseUnderstandingUserGoals2004, kellarFieldStudyCharacterizing2007}. As such,
closed-ended prompts asked participants what they would search to “find the
answer” whereas open-ended prompts asked participants what they would search to
“learn more.” Each of these sets of prompts includes one non-political prompt,
two political prompts related to abortion, and a final prompt asking the
participants to investigate a misinformed claim related to abortion. In total,
we gather eight Google search queries from each of the participants.

\begin{table}[]
  \resizebox{\columnwidth}{!}{%
  \begin{tabular}{@{}llll@{}}
  \toprule
  \textbf{Prompt}                                                                                                                                                                                                        & \textbf{\begin{tabular}[c]{@{}l@{}}Question\\ type\end{tabular}} & \textbf{\begin{tabular}[c]{@{}l@{}}Topic\\ category\end{tabular}} & \textbf{\begin{tabular}[c]{@{}l@{}}Example\\ query\end{tabular}}                                                     \\ \midrule
  \begin{tabular}[c]{@{}l@{}}... that asks how long you should roast a chicken.What \\ would you search in Google to find the answer?\end{tabular}                                                                       & Closed-ended                                                     & Non political                                                     & \textit{“How long to roast chicken”}                                                                                 \\
                                                                                                                                                                                                                         &                                                                  &                                                                   &                                                                                                                      \\
  \begin{tabular}[c]{@{}l@{}}... that asks whether a fetus can survive on its own outside\\ the womb 12 weeks after conception. What might you \\ search in Google to find the answer?\end{tabular}                      & Closed-ended                                                     & Political                                                         & \textit{\begin{tabular}[c]{@{}l@{}}“can a fetus survive outside the\\ womb on its own”\end{tabular}}                 \\
                                                                                                                                                                                                                         &                                                                  &                                                                   &                                                                                                                      \\
  \begin{tabular}[c]{@{}l@{}}... about the number of states that have outlawed abortion.\\ What would you search in Google to find the answer?\end{tabular}                                                              & Closed-ended                                                     & Political                                                         & \textit{\begin{tabular}[c]{@{}l@{}}“How many states have outlawed\\ abortion”\end{tabular}}                          \\
                                                                                                                                                                                                                         &                                                                  &                                                                   &                                                                                                                      \\
  \begin{tabular}[c]{@{}l@{}}... that asks whether the risk of breast cancer increases after\\ receiving an abortion. What would you search in Google to\\ find the answer?\end{tabular}                                 & Closed-ended                                                     & Misinformation                                                    & \textit{\begin{tabular}[c]{@{}l@{}}“can abortion cause breast cancer\\ in women?”\end{tabular}}                      \\
                                                                                                                                                                                                                         &                                                                  &                                                                   &                                                                                                                      \\
  \begin{tabular}[c]{@{}l@{}}... the benefits of washing chicken before you cook it. What\\ would you search in Google to learn more?\end{tabular}                                                                       & Open-ended                                                       & Non political                                                     & \textit{“is washing chicken useful”}                                                                                 \\
                                                                                                                                                                                                                         &                                                                  &                                                                   &                                                                                                                      \\
  \begin{tabular}[c]{@{}l@{}}... about a doctor who lost their license and got fired for\\ performing an abortion in a state where abortions were\\ illegal. What would you search in Google to learn more?\end{tabular} & Open-ended                                                       & Political                                                         & \textit{“doctor fired for abortions”}                                                                                \\
                                                                                                                                                                                                                         &                                                                  &                                                                   &                                                                                                                      \\
  \begin{tabular}[c]{@{}l@{}}... about a Catholic bishop who was forced to step down\\ because of his position on abortion. What would you\\ search in Google to learn more?\end{tabular}                                & Open-ended                                                       & Political                                                         & \textit{\begin{tabular}[c]{@{}l@{}}“catholic bishop steps down\\ abortion”\end{tabular}}                             \\
                                                                                                                                                                                                                         &                                                                  &                                                                   &                                                                                                                      \\
  \begin{tabular}[c]{@{}l@{}}... about an illegal sale of fetuses by Planned Parenthood.\\ What would you search on Google to learn more?\end{tabular}                                                                   & Open-ended                                                       & Misinformation                                                    & \textit{\begin{tabular}[c]{@{}l@{}}“is anyone selling fetuses illegally\\ through Planned Parenthood?”\end{tabular}} \\ \bottomrule
  \end{tabular}%
  }
  \caption{Participants are asked to formulate queries for each prompt
  open-ended and closed-ended. Each prompt starts with the phrase: "Imagine you
  see a post on social media...".}
  \label{tab:prompts}
\end{table}

To examine these queries computationally and understand how these queries were
formulated, we convert the queries into a more computationally measurable
representation. To achieve this, we deconstruct a query by extracting three
feature categories: vocabulary, style, and semantics. This allows us to
represent each query as three separate vectors, one for each feature category.
In the following section, we provide methods through which we convert the
queries into each of the feature vectors.

\para{Choice of Vocabulary.}
Words are the foundational elements of language without which people cannot
convey their intended meaning \cite{ghazalLearningVocabularyEFL2007}. This makes users' selection of
words a key component in understanding user behavior during information seeking.
These choices are often subconscious and influenced by a variety of factors
\cite{daelemansExplanationComputationalStylometry2013}. Indeed, prior work has repeatedly shown how our perceptions
shape how we use words --- e.g., our choices of vocabulary are influenced by our
communities \cite{senTaggingCommunitiesVocabulary}, past experiences \cite{paradisFrenchEnglishBilingualChildren2003}, and
even perspectives on specific issues \cite{beigmanklebanovVocabularyChoiceIndicator2010}.
Taken together, these reasons emphasize the importance of extracting and
encoding the vocabulary choices made by users during query construction for
information-seeking tasks. 

We encode a user's vocabulary selection during an information-seeking task by
assigning an “importance score” to each word in their query. We assign these
scores using a technique called TF-IDF (Term Frequency – Inverse Document
Frequency) after minimal preprocessing to remove punctuation and English stop
words. Simply put, TF-IDF assigns higher scores to words that are important
within a single query and simultaneously not common to many queries. We use this
approach to convert every user-generated query into a numerical vector. As shown
in the equation below, the magnitude of each dimension in this vector denotes
the importance of a word to the query, relative to all other queries. Here,
$I_q$ denotes the vector generated for a query $q$, w denotes a word from the
set of all words observed $W$, $q$ denotes a user-generated query from the set
of all user-generated queries $Q$, and $\textup{freq}(w, q)$ denotes the frequency
of the word $w$ in query $q$.

\[I_q[w \in W] = \textup{freq}(w, q) \cdot \log\frac{|Q|}{|\{q \in Q : w \in q\}|}\]

\para{Linguistic style.}
The ways in which people express identical ideas differ because individual
personality differences are reflected in our linguistic style
\cite{pennebakerLinguisticStylesLanguage2000}. Our extraction and encoding of
users' vocabulary selections in their queries are the first-order features
extracted from the manifestation of these differences. We also extract three
sets of second-order features which we expect to reflect the linguistic style
differences between users during query construction for information seeking. 

First, we use the LIWC (Linguistic Inquiry and Word Count) toolkit to extract
the psycholinguistic features of the user-created query. These features provide
a representation of the user's attentional focus, social relationships,
emotionality, thinking styles, and other elements of cognitive processing
\cite{tausczikPsychologicalMeaningWords2010}. Each feature represents one psycholinguistic
assessment (e.g., fear, negative sentiment, etc.) and is represented by a
numerical score between 0 and 1. These scores indicate the fraction of the words
in the query that contribute towards a specific psycholinguistic assessment.
Next, we include a set of ten features that capture the frequencies of ten
different parts of speech occurring in each user-generated query. These features
contribute to an assessment of the user's linguistic style during the
information-seeking task --- e.g., the higher prevalence of adjectives reflects
a more descriptive style. These features are generated using the part-of-speech
tagger from the NLTK toolkit. Finally, we also include a set of
features that reflect the complexity of the language used within the query.
These include the total number of words and average word length. 

\para{Contextual semantics.}
Our vocabulary selection and linguistic style features rely on syntax and are
context-agnostic. Put differently, each of the prior methods treats a query as a
“bag of words” during feature construction (i.e., words are considered
independently of the order or context in which they are used). Accordingly, we
also include features that consider the contextual semantics of a user-created
query to allow for a more complete understanding of the underlying concepts,
themes, and biases contained within it. 

We accomplish our goal of integrating context by creating embeddings that
captures the meaning of each user-constructed query. We use the FastText
\cite{mikolovAdvancesPreTrainingDistributed2018} word embeddings, constructed
from the 2018 Common Crawl data, to obtain vectors for each word in a query.
These vectors represent the semantics of each of the query. Similar to methods
used to compute bias within word embeddings,
\cite{caliskanSemanticsDerivedAutomatically2017} we compute the difference in
distance of these word vectors with the vectors associated with the two words:
“pro-life” and “pro-choice.” This indicates the implicit association that the
word has with the pro-life concept (if the difference is positive) or the
pro-choice concept (if the difference is negative). Finally, we compute the
weighted average of these differences (weighted by the “importance score” of
each word in the query) to obtain a scalar value that indicates the abortion
polarity of the query.

\subsection{Search Results}
\label{sec:data:search-results}
Finally, the participants are asked to perform Google searches using their
queries written for the 4 open-ended prompts. We ask them to share the search
results page they were presented with for each of the queries. The particpants
share the entire search result page, with at most 10 search results, by copy and 
pasting the entire contents of the page. These search results contain three key
data points: 1) the domain of the search result, 2) the title of the search
results, and 3) the snippet of text presented below the search results which
acts as a description of the result.

\para{Search result page.}
In our analysis, we seek to perform analysis on the search results to primarily
determine whether participants with different stances are presented with
different results. The search engine's algorithm can be broken down into two key 
components: retrieval and ranking. The retrieval algorithm determines the set of
results that are relevant to the query and the ranking algorithm determines the
order in which these results are presented. To determine the influence of stance
on the search results, we limit the scope of our study to investigating the
retrieval algorithm. To this end, we examine the domains, titles and snippets
presented on the page regardless of their ranking. We obtain the vocabulary and
semantic representation of the titles and snippets by using the same methods as
described in the previous section for the queries. Whereas for the domains, we
represent them as simple unique tokens. For additional context, we obtain the
partisan score for domains and the ideological score for the titles and snippets.

\para{Partisan scores.}
We include the partisan score for the domain obtained from Media Bias/Fact Check
(MBFC). MBFC provides domain-level scores on the bias and factual reporting for
media and news sources. MBFC uses a combination of qualitative and quantitative
evaluations of headlines and articles from outlets. Using the bias score, we
score each domain present in a search result page. In addition to the bias
score, we also include other MBFC metrics such as credibility score and factual
reporting score.

\para{Ideological score.}
The content of the search results and search queries, specifically their
semantics, encode ideological bias. To measure the ideological bias or score of
the search results -- specifically the titles and snippets -- we convert the 
text into embedding representation using word embedding models, in our case
FastText. These embeddings represent our search results in a latent space where
words that are found frequently together or are used in similar contexts in the
word embedding model's training dataset (i.e., the common crawl) are likely to
be given embeddings that are closer together in the latent space. For example,
the words pro-abortion and pro-choice are likely to occur together would
therefore be given embeddings that are closer together. To measure the
association or bias of the search results with the pro-life or pro-choice
ideology, we compute the distance of the embedding from the words “pro-life” and
“pro-choice.” The difference between these distances is considered the pro-life
association score which represents the association of the search result or the
search query with the pro-life ideology. This method of measuring the
ideological bias of text is similar to the methods used to compute bias within
word embeddings \cite{caliskanSemanticsDerivedAutomatically2017}.
\section{Do individuals with opposing attitudes receive different results?}
\label{sec:rq1}
We first investigate the extent to which individuals with opposing attitudes on
the topic of abortion receive different search results. Significantly different
search results would suggest the individual's preexisting beliefs are in fact
factored into the information-seeking process. We group together participants on
the same side of our abortion scale into pro-life and pro-choice users. We find
participants with opposing attitudes are presented with significantly different
search results.

\para{Comparing search results.}
Once participants with similar attitudes towards abortion are put in groups, we
compare whether the domains and the information present in the result's title
and snippet are significantly different across these groups. To this end, for
each result aspect (i.e., domain, title, and snippet) we perform a permutation
test that compares the difference across these groups with the difference across
random permutations of groups. The permutation test compares the difference
between the two distributions with the differences between random permutations
of the distributions together. This process generates a distribution of test
statistics under the null hypothesis of no group difference. The $p$-value is
determined by the proportion of permuted test statistics that are greater than
the observed statistic, thereby providing an exact significance level without
making strong parametric assumptions about the underlying data distribution.  We
determine a statistically different distribution of characteristics between the
two groups by comparing their differences with the differences obtained from
10,000 randomly permuted groups. If the difference measured between our two
groups is larger than those observed in 9,500 of the random permutations, we
claim that the two groups have statistically significant differences (since $p$ <
.05).

\begin{figure}[!htbp]
  \centering
  \includegraphics[width=\textwidth]{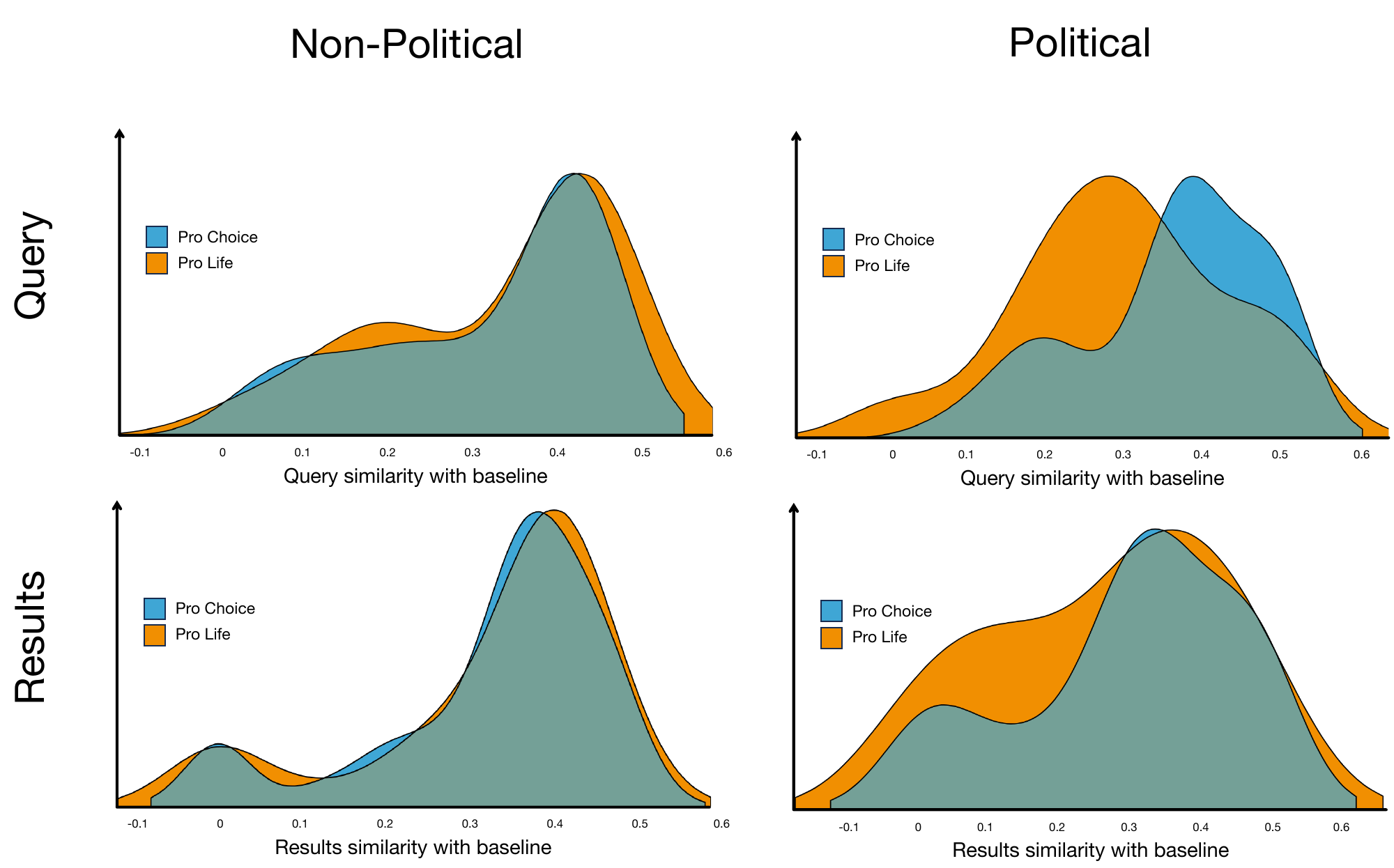}
  \caption{KDE distribution of similarity of queries and results with the
  neutral baseline. Search queries and results for political prompts, as opposed
  to non-political prompts, have dissimilar similarities with baseline across
  pro-life and pro-choice users.}
  \label{fig:similarity_differences}
\end{figure}

\para{People with opposing attitudes are presented with different information.}
Comparing the results from pro-life and pro-choice participants using a
permutation test, we find the dissimilarity to be significantly greater, by at
least 16\% (p < 0.05), than the dissimilarity between randomly permutated groups of
participants. In other words, the titles, snippets, and sources returned to
pro-life and pro-choice users significantly differed from each other (as also
demonstrated in \Cref{fig:similarity_differences}). Notably, we find no
difference between search results for pro-life and pro-choice users for the
non-political prompts. Simply put, an individual's preexisting beliefs
influence their search results. This finding suggests that political beliefs are
communicated somehow during the information-seeking process. These findings are
consistent with prior experiments
\cite{leMeasuringPoliticalPersonalization2019a} showing that a user's browsing
history affects their search results. Yet, while their study relied on automated
profiles trained to visit political content, ours incorporates actual users who
hold measurable political views. Our findings further beg the question of
whether the individual's preexisting beliefs are revealed to the
information-seeking process (specifically Google) through explicit sources (i.e.
information explicitly communicated by the user) isolated to a person's query
\cite{slechtenAdaptingSelectiveExposure2022,vanhoofSearchingDifferentlyHow2022},
or implicit sources such as their search history.

\section{Are preexisting beliefs encoded within queries, influencing the search results?}
\label{sec:rq2}
Online platforms provide users with communication channels through which their
information-seeking tasks and identity characteristics, such as their political
attitudes, are communicated to the platform. These modalities can range from
explicit to implicit. Explicit modalities involve users actively and directly
communicating with the platform, for example, when users communicate their
information task through a query. Implicit modalities, on the other hand,
indirectly reveal user information such as their identity characteristics, for
example, revealing user's political attitudes through search history. Here, we
examine how each of these modalities influences search results. Specifically, we
investigate whether characteristics of the query i.e., the choice of vocabulary,
writing style, and the semantics of the query, reveal unintended implicit
characteristics of the user. These user characteristics may include their
preexisting attitudes towards the topic of the query. We find participants with
opposing attitudes formulate queries differently, crucially, using different
choices of words. This nuanced difference in vocabulary subsequently influences
the difference in search results.

\para{Comparing search queries.}
To establish whether people with different preexisting beliefs on a topic
formulate queries differently, we compare queries formulated by participants
with opposing attitudes. For each query, we create vectors that represent the
query's choice of vocabulary, writing style, and semantics of query (see
\Cref{sec:data:queries}). We group participants on the same side of our abortion
and compare these characteristics of queries across groups with opposing
attitudes using a permutation test. Observing a significant
difference in choice of vocabulary or writing style would suggest that there is
an unintended implicit difference in queries when people seek information on the
topic, they have opposing attitudes on, whereas a significant difference in the
semantic of the query, i.e. the meaning of queries or the information-seeking
task, could suggest people seeking different information for the task at hand --
calling for further analysis.

\begin{figure}[!htbp]
  \centering
  \includegraphics[width=\textwidth]{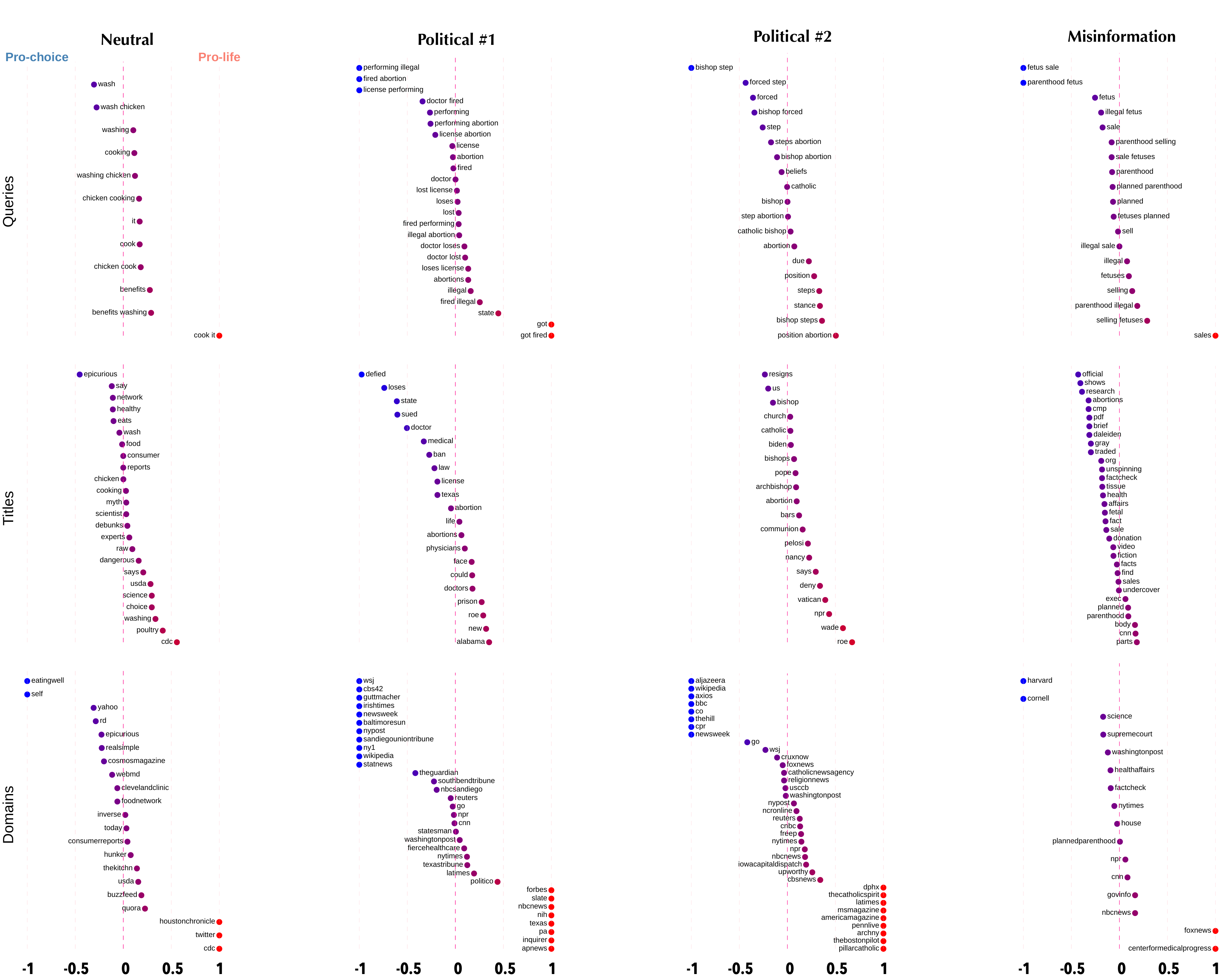}
  \caption{Comparing query language, result titles, and domains between pro-life
  and pro-choice groups.
  \newline
    \emph{Vocabulary frequency in queries:} This chart shows the vocabulary and phrases (unigrams and bigrams) used in search queries by both pro-life and pro-choice participants. The value scale ranges from -1 to 1, indicating how likely a term is to be used by one group over the other. A positive score means the term is more frequently used by pro-life participants, with 1 signifying exclusive use by this group, and vice-versa. Only vocabulary and phrases found in more than 5
  \newline
    \emph{Vocabulary in search result titles:} This part of the figure shows the frequency of specific words appearing in the titles of search results shown to participants from both groups.
  \newline
    \emph{Domain frequency in search results:} This section identifies the frequency at which different website domains appear in the search results provided to both pro-life and pro-choice participants. 
  }
  \label{fig:vocabulary}
\end{figure}

\para{Participants with opposing attitudes formulate queries differently.}
We find that participants with opposing attitudes formulated queries with
similar semantics (e.g., the meaning of the query) and writing styles (e.g., the
linguistic feature); however, the choice of vocabulary (e.g., the actual words
used) in the queries was significantly different between pro-life and pro-choice
users. The vocabulary that participants use in formulating their queries,
specifically when performing information-gathering searches in response to
open-ended prompts, is significantly regulated by their preexisting attitudes.
Using our permutation test, we find the difference in vocabulary between
pro-life and pro-choice participants to be significantly greater (p > 0.05)
compared with the difference between randomly permutated groups of participants.
The dissimilarity of vocabulary between participants with opposing attitudes
considering the semantic similarity of their queries highlights the implicit
differences in queries formulated by each group. We demonstrate this difference
in vocabulary in \Cref{fig:vocabulary} with a specific focus on bi-gram vocabulary. The
vocabulary used to express their information-seeking task separates participants
with opposing attitudes. In other words, pro-life and pro-choice users formulate
search queries that are similar in meaning yet distinct in word choice. We
hypothesize that this distinction enables collaborative filtering participants
with similar attitudes on abortion to be grouped, which explains different
search results. Regardless, these findings provide evidence that users'
preexisting attitudes are communicated through their choice of vocabulary. To
investigate whether these differences in vocabulary influence search results we
perform a mediation analysis.

\begin{figure}[!htbp]
  \centering
  \includegraphics[width=\textwidth]{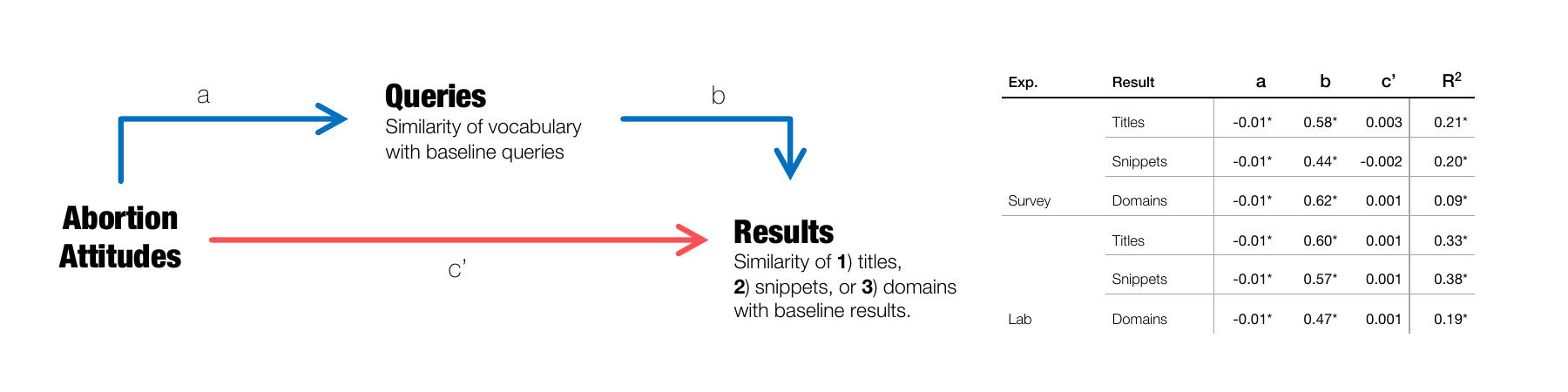}
  \caption{Mediation analysis examining the influence of abortion attitudes on
  results through queries for the political prompt. The table presents mediation
  results for the first political prompt. In our mediation model, the variable
  “a” represents the effect of abortion attitudes on the vocabulary of queries,
  “b” represents the effect of the vocabulary of queries on results, and "c'"
  denotes the direct effect of attitudes on results. The R2 value quantifies the
  extent to which the variance in the results can be explained by the attitudes,
  queries, and their interaction. * Indicates statistically significant results
  ($p$ < 0.05).}
  \label{fig:mediation}
\end{figure}

\para{Modeling queries as a mediator between user attitude and search results.}
We design a mediation model of influence to investigate whether the search
results are influenced by participants' preexisting attitudes through the
queries that encode them. The attitudes toward abortion function as the
independent variable, implicit characteristics within the queries (the
vocabulary choice) as the mediator, and characteristics of the results (e.g.,
the domains, titles, and snippets present in the search results) as the
dependent variables, as shown in \Cref{fig:mediation}. We consider the queries formulated by
participants with neutral attitudes toward abortion as baseline queries and
results presented to participants with neutral attitudes toward abortion as the
baseline results. Using this baseline, we represent the queries as the deviation
of vocabulary from the baseline and the results as the deviation of the titles,
snippets, and sources from the baseline results. This mediation model,
therefore, measures the effect of participant's abortion attitudes on the
deviation of results from the baseline through its effect on the deviation of
queries from the baseline. Given the multiple components of search results
(titles, snippets, and domains) and search queries (vocabulary, style, domains),
we aim to investigate the influence of each representation separately.

\para{Participant's attitudes influence results through the implicitly encoded queries.}
Our mediation model provided two key results. 1) We do not observe any
significant direct effect of attitude towards abortion on the search results;
however, 2) attitude towards abortion has a significant indirect effect on the
results through its effect on queries. More specifically, taking a political
prompt as an example, we show, in \Cref{fig:mediation}, the significant indirect effect
abortion attitudes have on the titles, snippets, and domains presented to the
user through its influence on the vocabulary of queries.

These results, combined with our earlier finding that participants with
different attitudes are presented with different results, suggest that the
participant's attitude-encoded query influences the search engine to present
different search results to participants based on their attitude towards
abortion. This interpretation is further supported by our model being able to
significantly explain the variance of the search results (0.3 R2). Building
different models for each prompt of varying levels of political nature (a
non-political, two political, and a misinformation prompts), we observe the
indirect effects of stance on results to remain significant. However, we
observed a stronger effect of queries on the divergence of results on political
and misinformation prompts. These results imply that the influence of a user's
stance on abortion on results is greater for users seeking political or
polarizing information. 

\para{Collaborative filtering.}
One explanation for how vocabulary reveals the participants' attitudes toward
abortion includes the process of collaborative filtering. To optimize user
engagement, search engines often rely on collaborative filtering, which presents
relevant results to users based on the engagement patterns of previous users
\cite{RankingResultsHow}. Essentially, collaborative filtering is a technique in which inputs of
prior users and their subsequent engagement with the suggested content are used
in measuring the relevancy of results with queries \cite{ricciIntroductionRecommenderSystems2011}. In other words, the
engagement patterns of prior users can influence the results of subsequent
searches with similar queries. This collaborative approach to contextualizing
the relevance of information adds a new dimension of content interpretation to
the information-seeking process. We learn from our prior analysis that users
with similar attitudes formulate queries with similar choices of vocabulary.
Additionally, users might disproportionately engage with information based on
their preexisting beliefs \cite{slechtenAdaptingSelectiveExposure2022,
robertsonUsersChooseEngage2023}. Therefore, as individuals write queries with
vocabulary that reflect their preexisting beliefs, they might be presented with
results that are engaging to users who used similar vocabulary (i.e., users with
similar preexisting beliefs).

Taken together, we speculate the interaction of collaborative filtering and
cognitive biases to yield equivalent results for participants with similar
attitudes and different results for participants with opposing attitudes. As our
results show, pro-choice users tend to enter search terms that are similar to
those of other pro-choice users. In turn, they receive search results that have
been engaged with by earlier pro-choice users. This creates a theoretical
feedback loop that further reinforces alternate information-seeking outcomes and
realities

\section{Does search history contribute to the divergence of results?}
\label{sec:rq3}
Search engines include additional sources of information to learn additional
identity characteristics of their users to optimize search quality. These are
more implicit channels that communicate characteristics, such as location,
device type, and a user's search history. These supplementary sources of
information allow the platform to build more faithful perceptions of its users
that may result in a more relevant user experience. As indicated in section \Cref{sec:data:queries}
we prompt our participants to formulate queries for various information tasks
and perform the search on their computers -- which include their personalization
and search history. In our analysis, the search results we analyzed so far
included the influence of search history and personalization. In this section,
we investigate the extent to which search history and personalization influence
the divergence in search results for participants with opposing attitudes. 

\para{Removing search history from information tasks.}
We examine this by repeating our analysis of search results with the influence
of search history and personalization removed. We remove this influence by
performing searches for the four open-ended prompts using queries formulated by
participants in a controlled lab environment. In the controlled lab environment,
we perform the information tasks on new blank user profiles with no search
history. Doing so ensured that we eliminated the influence of search history and
user personalization on the search process. Subsequently, we repeat our
mediation analysis to investigate whether the influence remains significant. The
only difference in the mediation analysis is in search results, which now lack
the influence of search history and personalization.

\para{User personalization and search history influence the information-seeking process.}
Similar to the prior model, this mediation model measures the influence of the
participants' preexisting attitudes on their search results through their
formulated queries, however, without any personalization or search history
influencing the search results. Our model yields results of similar significance
and magnitude as our prior model with in-the-wild search results. However, in
comparing the explainability of search results, we observe a 45
values (from 0.5 R2 to 0.3 R2) when search history and other implicit modalities
are not included. Since removing the personalization in the lab experiment
reduced the explainability of search results, we explain that the presence of
personalization and search history are invisible and untested mediators in our
prior models.

Taken together, we find user characteristics are revealed through passive
sources (i.e., search history and personalization) and active yet implicit
sources (i.e., vocabulary differences). The two sources highlight the control or
lack thereof a user has in seeking neutral information that is unbiased through
their preexisting beliefs. 
\section{Google serves results to reinforce preexisting beliefs.}
\label{sec:rq4}
Our findings, so far, have established the mechanisms through which the
information-seeking process involves the user's preexisting attitudes towards
the topic and provides different results based on their attitudes. We interpret
these findings to explain how the role of collaborative filtering, a feature of
modern information-seeking processes, amplifies the human cognitive confirmation
bias and serves similar content to individuals with similar characteristics.
Finally, we test whether the information-seeking process yields ideologically
congruent information to construct a filter bubble effect. Based on our findings
above, we hypothesize individuals are served with content that reinforces their
preexisting beliefs. To this end, we measure whether participants receive 1)
search results from sources that correspond with their attitudes and 2) search
results containing information that is congruent with their attitudes. This
section therefore investigates whether the Google Search information-seeking
process creates a filter bubble effect where its users are shown
disproportionate results that reinforce their preexisting beliefs.

\para{Google constructs an epistemic bubble based on the user's preexisting beliefs.}
We measure whether the information-seeking process constructs epistemic bubbles
\cite{nguyenECHOCHAMBERSEPISTEMIC2020} -- where participants are served results that are ideologically congruent
with their preexisting attitudes. Our analysis, using the partisan scores of 
the sources (see \Cref{sec:data:search-results}), reveals participants with
pro-life attitudes are significantly more likely to be presented with results
from right-wing sources than left-wing sources, while pro-choice participants
were significantly more likely to be presented with results from left-wing
sources than right-wing sources. This correlation between abortion attitudes and
political partisanship (i.e., pro-life association with conservative
\cite{centerAmericaAbortionQuandary2022, incAbortionTrendsParty2018} and
pro-choice association with liberal) is corroborated by prior work as well as
the significant correlation between the political learning and abortion
attitudes of our participants. More notably, we identified a disproportionate
distribution of low-credibility sources among participants with pro-life
attitudes toward abortion when searching for the misinformation prompt. Removing
the search history and personalization signals to the information-seeking
process, we observe the correlation between abortion attitudes and the political
affiliation of sources to dampen. Furthermore, the observed significant positive
correlation between low-credibility sources and pro-life attitudes disappears.
These findings suggest that identity characteristics of users revealed through
personalization and search history modulate the sources of information more than
the process of formulation of queries.

\para{Measuring ideological score.}
For all participants, we computed the ideological scores of their search queries and the
search results they were presented. Following this, we measure the distances of
these embeddings with the keywords “pro-life” and “pro-choice”, as described in
\Cref{sec:data:search-results}. The difference between these distances is
considered the pro-life association score which represents the association of
the search result or the search query with the pro-life ideology. Finding a
significant correlation between the ideological score of the search results and
the abortion attitude of the participant would suggest the presence of
ideological congruency within the information-seeking process whereas the
presence of ideological congruency between queries and attitudes would indicate
the presence of cognitive biases.

\para{The modern information-seeking process creates a filter bubble effect.}
Finally, through semantic interpretation of the titles and snippets, we
identify the presence of ideological congruency between the search results and
participants' preexisting ideologies. We semantically interpret the titles and
snippets of the search results and the queries to measure their similarity with
both attitudes towards abortion. Our
investigation yields a significant and positive correlation between their
pre-existing attitudes and the semantic attitudes towards abortion present in
search results. However, we find no significant ideological congruency present
between the participants' stances and their queries. This suggests participants
were presented with search results that supported and reinforced their
preexisting beliefs regardless of the semantic presence of the attitudes in the
queries. Since queries were not semantically associated with participants'
preexisting attitudes on abortion, possible sources for the ideological
congruency found in the results include a combination of search history and
choice of vocabulary within the queries.	

\section{Discussion} 
\label{sec:discussion}
The modalities through which we seek information has changed significantly since
the advent of the Web. Algorithms dominate as mediators to information be it
news, instructional, entertainment, or educational. In this work, we study how
the modern information seeking processes is vulnerable to creating a filter
bubble effect, a problematic pattern dangerous for democratic societies. We
focus our study on the process of seeking information through Google Search and
investigating the outcomes and sources of biases through preexisting attitudes
of a user. 

\subsection{Limitations}
Given these findings and conclusions, it is important to recognize that our
work's takeaways are limited by the representation and diversity of participants
and its focus on a single issue for the information task. The 226 participants
for the study were primarily composed of undergraduates from a university on the
East Coast. The demographic represented by our participants may not be
representative of the broader population and therefore the findings may not be
directly applicable. However, since our work corroborates findings from prior
works, we do not foresee unexpected conclusions from different demographics
within the US. Moreover, the selection of abortion for the topic of focus was
driven by the contentious debate surrounding the topic during the overturning of
Roe v. Wade by the supreme court. During this time, there was a considerable
increase in the volume of searches on Google for information related to
abortion, as demonstrated by the Google Search Trends \footnote{https://trends.google.com/trends/explore?date=today\%205-y\&geo=US\&q=abortion\&hl=en}. We sought to
simulate real world information seeking behavior during the time of contention
on a topic and therefore formulated real world scenarios for participants to
engage. However, due to the special circumstances of the topic of
abortion, we acknowledge the possible limitation of our findings when
generalizing to less contentious, established, or neutral topics. Despite this
limitation, we consider filter bubbles effect for contentious and trending
political topics to be problematic, even more so, than regular topics. Finally,
methodologically, utilizing manual coding by experts as compared to
computational methods used in our work would have provided a more grounded and
validated results, however, the resources required for such an undertaking were
prohibitively expensive. Therefore, the methods we use to represent queries and
search results computationally are “best-effort” proxies that have been
validated comprehensively by prior works. 

\subsection{Conclusions and implications}
The key takeaway of our work is the personalization of search results by Google
Search based on user's preexisting attitudes. We do this by recruiting
participants with different attitude towards abortion and prompting them,
through a range of realistic scenarios, to search for information related to
abortion on Google. Our study demonstrates, participants with opposing attitudes
on abortion, when searching for information related to abortion, were shown
search results with meaningful and significant differences. The search results
were influenced by the participant's preexisting attitudes, specifically the
sources of the results had similar bias as the participant's attitude towards
abortion, and the results' titles and snippets had ideological congruency with
the participant's attitude. Even with search history and personalization
controlled for, Google Search yields different search results for participants
across groups. We demonstrate the significant difference in the information
seeking process for participants from across the groups to be in the choice of
words when formulating the queries. The differences in the vocabulary of the
queries across groups were significant enough to separate the participants into
the groups by their choice of words alone. We conclude, through our mediation
analysis, this difference in vocabulary to be a significant mediator in
communicating the user's preexisting attitudes to Google Search resulting in
personalized search results. Crucially, this uncovers the alarming power of
unintentional implicit signals in our language that play a key role in this
phenomenon. These encoded preexisting beliefs, specifically within the language
of the query, when combined with the speculated interaction of collaborative
filtering and user cognitive bias, present users with ideologically congruent
information. In some cases, when considering implicit forms of information, it
also resulted in certain users (i.e., those with pro-life attitudes) receiving
information from sources lacking in credibility. This research demonstrates the
existence, causes, and consequences of algorithmic personalization in the search
process, thereby providing a relatively comprehensive depiction of the process
of information seeking in the modern information environment.
\balance


\bibliographystyle{ACM-Reference-Format}
\bibliography{google}



\end{document}